# Preferential arrangement of uniform Mn nanodots on Si(111)-7×7 surface

De-yong Wang, Li-jun Chen, Wei He, Qing-feng Zhan and Zhao-hua Cheng*

*State Key Laboratory of Magnetism and International Center for Quantum Structures, Institute of Physics, Chinese Academy of Sciences, Beijing 100080, P.R. China*

## Abstract

Under proper growth conditions, ordered and uniform Mn nanodots were fabricated on the Si(111)-**7×7** surface without the presence of a wetting layer. Furthermore, the Mn nanodots deposited onto the elevated substrates were observed to occupy preferentially on the faulted half unit cells (FHUCs) of the Si(111)-7×7 surface. This phenomenon implies that the Mn dots adsorbed on the FHUCs is more stable than those adsorbed on the unfaulted half unit cells (UFHUCs). Within the framework of quasiequilibrium thermodynamics, the energy difference between adsorption on the UFHUCs and the FHUCs was estimated to be 0.05±0.01 eV. The intrinsic attractive potential wells on the FHUCs effectively trap the outdiffusion of Mn atoms, and consequently result in a preferential arrangement of islands with well-defined sizes.

PACS numbers: 61.46.+w, 68.35.Bs, 75.75.+a

*Corresponding author

E-mail:zhcheng@aphy.iphy.ac.cn

Artificially controllable fabrication of nanostructures on supporting substrates with atomic precision has long been a subject of fundamental and applied science research [1-5]. Recently, ordered and identical Indium, Gallium, Aluminum and Cobalt nanoclusters have been fabricated by using silicon surfaces as templates [6-9]. Although the uniform magnetic clusters can be grown on insulating substrates[10,11], up to date, highly ordered and uniform ferro-, ferri, and antiferromagnetic nanodots on the supporting Si(111) substrates have not yet been achieved.

In this work, the Si(111)-**7×7** reconstructed surface was employed as a template to grow monodispersed Mn nanodots. The reasons for choosing Mn deposition on the Si(111)-**7×7** are followings: (1)Mn is a unique element in the $3d$ transition metal series and small Mn clusters are expected to possess anomalous magnetic properties [12]; (2)The uniform nanodots grown on semiconducting substrates without a wetting layer are often desirable for electronic and magnetic device applications. It was found that three dimensional (3D) Mn islands were immediately formed on the Si(111) surface without the presence of a wetting layer(the Volmer-Weber (VW) growth mode)[13]. Furthermore, the periodic pattern of the reconstructed Si(111)-7×7 lattice has the potential to confine the Mn clusters growth and results in a uniform and ordered array of clusters. (3)Comparing with other magnetic metals, such as Fe [14], and Co [9] on Si(111), Manganese silicates are not easy to form at room temperature, even at elevated temperatures. Therefore, it provides a unique opportunity for investigating the low dimensional magnetism of metallic nanodots. By delicate control of the growth kinetics, ordered and uniform Mn nanodots on the Si(111)-**7×7** reconstructed surface were successfully fabricated.

The experiments were performed with a combined molecular beam epitaxy

(MBE)/scanning tunneling microscope (STM) system in ultrahigh vacuum (UHV) at a base pressure of about $1\times10^{-10}$ mbar. A Si(111) substrate was cleaned by resistive flashing in UHV until a high quality **7×7** reconstruction was observed by OMICRON variable temperature STM. High purity Mn (purity 99.99%) was heated at 650°C by a boron nitride crucible, and then deposited onto the reconstructed surface with a deposition rate of 0.167 monolayer (ML)/min(1ML=7.88 × $10^{14}$ atoms/cm$^2$). All room-temperature scanning tunneling microscopy (STM) images reported here were recorded with tunneling current ranging form 20 to 40 pA. A chemically etched tungsten tip was used as the STM probe.

When Mn atoms were deposited onto the Si(111)-7×7 reconstructed surface at room temperature with very low coverage(0.02 ML), they aggregated immediately into three dimensional (3D) islands coexisting with bare Si(111)-7×7 substrate, indicating the VW growth mode. The irregular Mn clusters (bright areas) distribute randomly on the faulted and unfaulted halves of Si(111)-7×7 unit cells, as indicated in Fig. 1. With increasing the coverage, both the shape and the size of Mn clusters become more irregularly [13]. The irregularity in sizes and shapes of Mn clusters originates from the statistical nature in the deposition and diffusion process. It was well-known that the fault half unit cells (FHUCs) of the Si(111)-7×7 are more reactive than the unfaulted half unit cells (UFHUCs) and the intrinsic attractive potential wells on the FHUCs effectively trap diffusing metal atoms [15]. Mn atoms are therefore expected to occupy preferentially on the FHUCs. However, the fact of the random distribution of Mn clusters on the Si(111) surface implies that some of Mn atoms aggregate into sizable clusters prior to reaching the stable position due to the low diffusion.

In order to increase the diffusion of Mn atoms and improve the spatial ordering of Mn nanodots, the substrates were heated at various temperatures. Figs. 2(a), 2(b), and 2(c) illustrates the STM images of the Mn clusters deposited on the elevated substrates with different substrate temperaturesare of 27°C, 120 °C and 180 °C, respectively. The coverage of Mn nanodots is about 0.07 ML. Images were obtained with the sample bias of -2.09V and tunneling current of 0.28nA. It can be seen that the deposition of Mn on the Si(111) does not disturb the **7×7** periodicity of Si(111) reconstructed surface. The continuing presence of the Si(111)-7×7 reconstructed surface indicates that the Mn atoms have not substantially reacted with the surface Si atoms even at 180 °C. With increasing the substrate temperature from room temperature to 180°C, a preferential occupancy of Mn dots on the FHUCs of the Si(111)-7×7 was observed by STM. This observation reveals that some Mn atoms deposited onto the elevated substrates can overcome the repulsive barrier and move cross the boundaries between the two halves of Si(111)-7×7 unit cell to form regular and stable dots with preferential occupancy. Assuming that the island formation process is controlled thermodynamically by a quasi-equilibrium situation [16], the energy difference of Mn adsorption between the two halves of the (7×7) unit cell can be estimated by applying the Boltzmann distribution

$$N_F/N_U = \exp(-\triangle E/k_B T)$$

where $N_F$ and $N_U$ are the occupation numbers on the FHUCs and UFHUCs, respectively. $T$ is the substrate temperature, $\triangle E$ is the energy difference between the two halves of the 7×7 unit cell, and $k_B$ is the Boltzmann constant.

On the basis of the occupancy numbers obtained from the STM images at $T$=180 °C, the energy difference $\triangle E$=0.05±0.01eV was yielded. This energy difference is comparable with those for Pb(0.05 eV) [17], Tl(0.075±0.01 eV) [16], and In(0.08±0.01 eV) [8] on the Si(111)- 7×7 reconstructed surface. Thus, from energy point of view, the Mn dots favor to stand on the FHUCs.

The effect of substrate temperature on the shape, diameter and height of the Mn nanodots is clearly illustrated in figures 2(d)-(i). It can been observed that the shape of Mn nanodot on the substrate temperature of 27°C is very irregular although the size distribution is relatively narrow. The size and shape of Mn nanodots are found to become more uniform and depend sensitively upon the substrate temperature.

Figure 3 (a) shows a representative STM image of the Mn dots with a coverage of 0.21 ML. The uniform nanodots distribute mostly on the FHUCs of the Si(111)-7×7 surface. Figure 3(b) is a close-up together with a line profile of the dots. Estimated from the line scan, the corresponding histograms of diameter and height of the nanodots are illustrated in Figures 3(c) and 3(d), respectively. The mean dot height, $<h>=0.24$ nm, and diameter $<d>=1.35$ nm, are derived by fitting the histograms with Gaussian function. The dispersions of height, $\triangle h=(<h^2>-<h>^2)^{1/2}$, and of diameter, $\triangle d= (<d^2>-<d>^2)^{1/2}$, are only 0.025 nm and 0.15 nm, respectively, which are about 10% of the average height and diameter values. Such narrow dispersions are particularly striking because dots grown via the VW mode normally have much broader distributions [1]. Although individual adatom is highly mobile within the potential wells, the outdiffusion from the potential wells is restricted by the boundaries between the FHUCs and the UFHUCs of Si(111)-7×7 reconstructed surface, and consequently well-defined size dots are formed owing to the quantum confinement effect.

Another striking feature for the uniform Mn dots is that their sizes are almost invariable with increasing the Mn coverage. For example, figure 4 illustrates the STM images of Mn dots at the coverages of 0.08 ML, 0.18 ML and 0.22 ML, together with the diameter, height and density of supported Mn dots. The occurrence of clusters coalescence is not observed by

STM even at the coverage of 0.22 ML. As can be seen in the STM images, the average nearest-neighbor distance between dots clearly decreases with increasing the coverage, while the average size is invariable. Furthermore, the density of Mn dots increases linearly with increasing the coverage. This is drastically different from the classical nucleation and growth model where a linear relationship between the dot density and adatom coverage is only valid at the beginning of nucleation stage (<<0.1 ML). Before the coalescence regime the variation of the average dot diameter with deposition coverage can be typically expressed by a power law relationship [18]. No correlation between the average dot size and the average nearest-neighbor distance of inter-dots is observed. This fact implies that, compared with the attractive potential wells on the FHUCs and the repulsive barriers at boundaries of two halves of unit cells, dot-dot dipolar interaction plays a minor role in the formation of Mn nanodots on Si(111)-7×7 surface.

In summary, the nanometer-scaled Mn dots with ordered and narrow size distribution were successfully fabricated on the Si(111)-7×7 reconstructed surface. It has been demonstrated that the vertical and lateral sizes of the dots is almost invariable with increasing the coverage of Mn nanodots. The preferential occupancy of Mn nanodots originates from the energy difference between the UFHUCs and the FHUCs of Si(111)-7×7 surface. The quantum confinement effect of attractive potential wells on the FHUCs results in the narrow size distribution of Mn nanodots.

This work was supported by the State Key Project of Fundamental Research, and the National Natural Sciences Foundation of China.

**FIGURE CAPTIONS**

FIG. 1 (color online). STM image (20nm×20nm) of Mn deposited on the Si(111)-7×7 surface at room temperature with a coverage of 0.02ML. Sample bias voltage $V_s$ = +1.93V.

FIG. 2(color online). (a), (b), and (c) are STM images (30×30 nm$^2$) of Mn nanodots grown on Si(111)-7×7 at room temperature, 120 °C and 180 °C, respectively. Images were obtained with the sample bias of -2.09V and tunneling current of 0.28nA. (d)-(i) Diameter and height distributions of the dots and the corresponding Gaussian fits.

FIG. 3 (color online). Nanoscale Mn dots grown on Si(111)-7×7 with a nominal Mn dose of 0.21 ML. (a) A typical STM image(30×30nm$^2$). (b) Close-up (20×20nm$^2$) and line profile (about 17nm long) of the dots. (c), (d) Diameter and height distributions of the dots shown in (a) and the corresponding Gaussian fits.

FIG. 4(color online). (a), (b), and (c) are STM images of Mn dots deposited on Si(111)-7×7 at Mn coverages of 0.08ML, 0.18ML and 0.22 ML, respectively. The sample bias and tunneling current are -2.0V and 0.27nA; (d), (e), (f) are the diameter, height and areal density of Mn nanodots as a function of coverage.

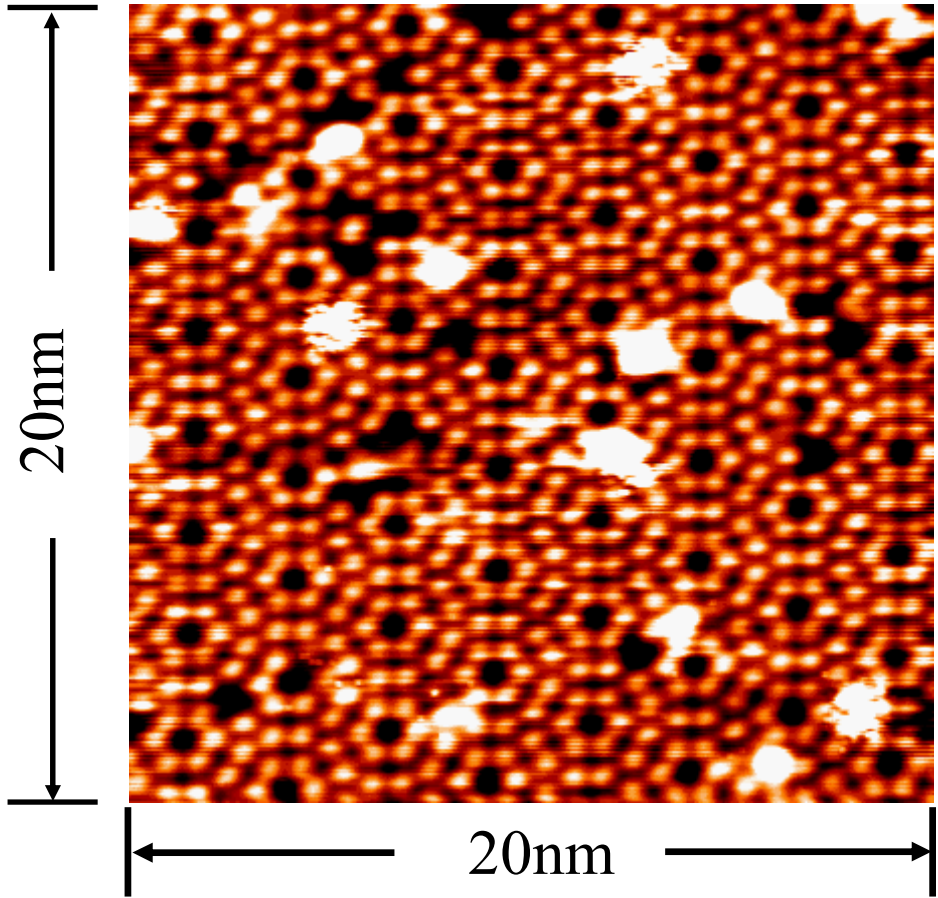

FIG.1

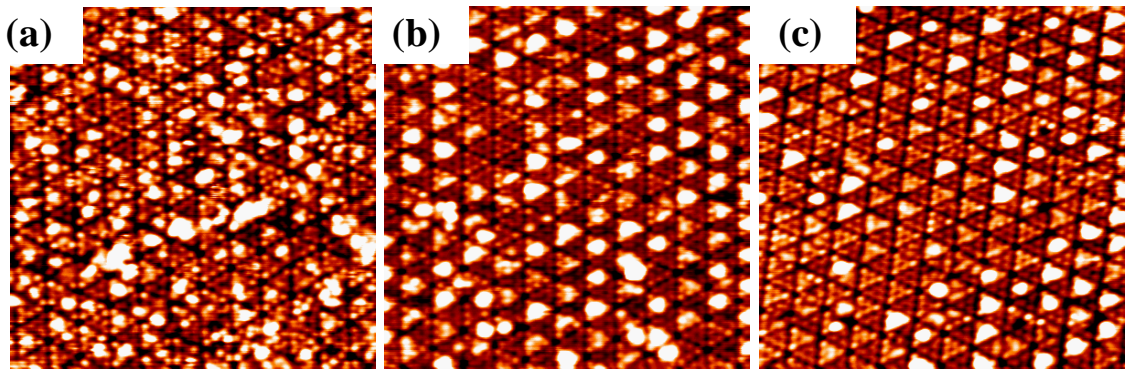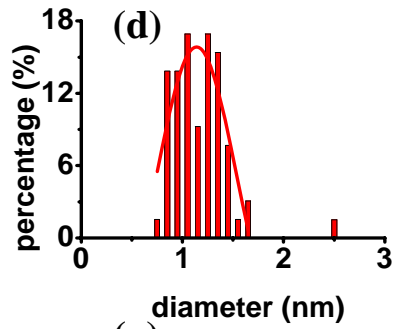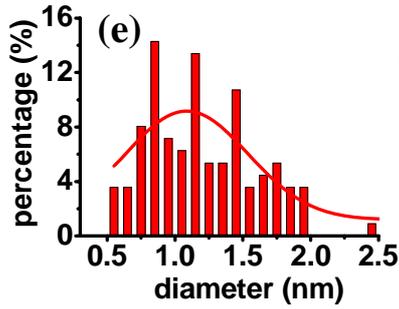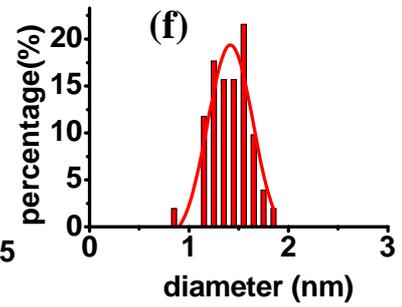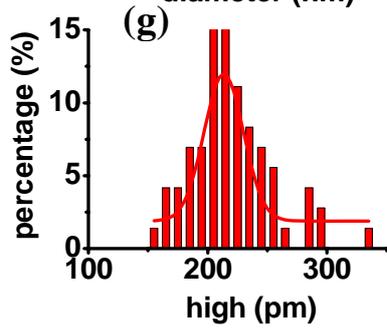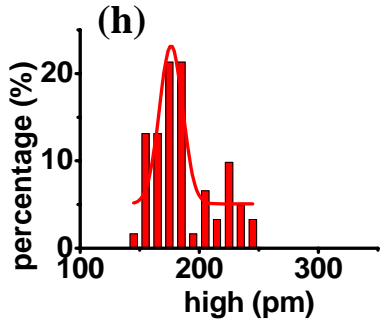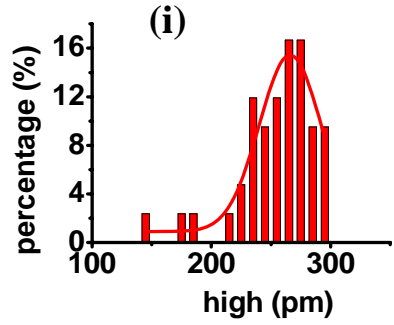

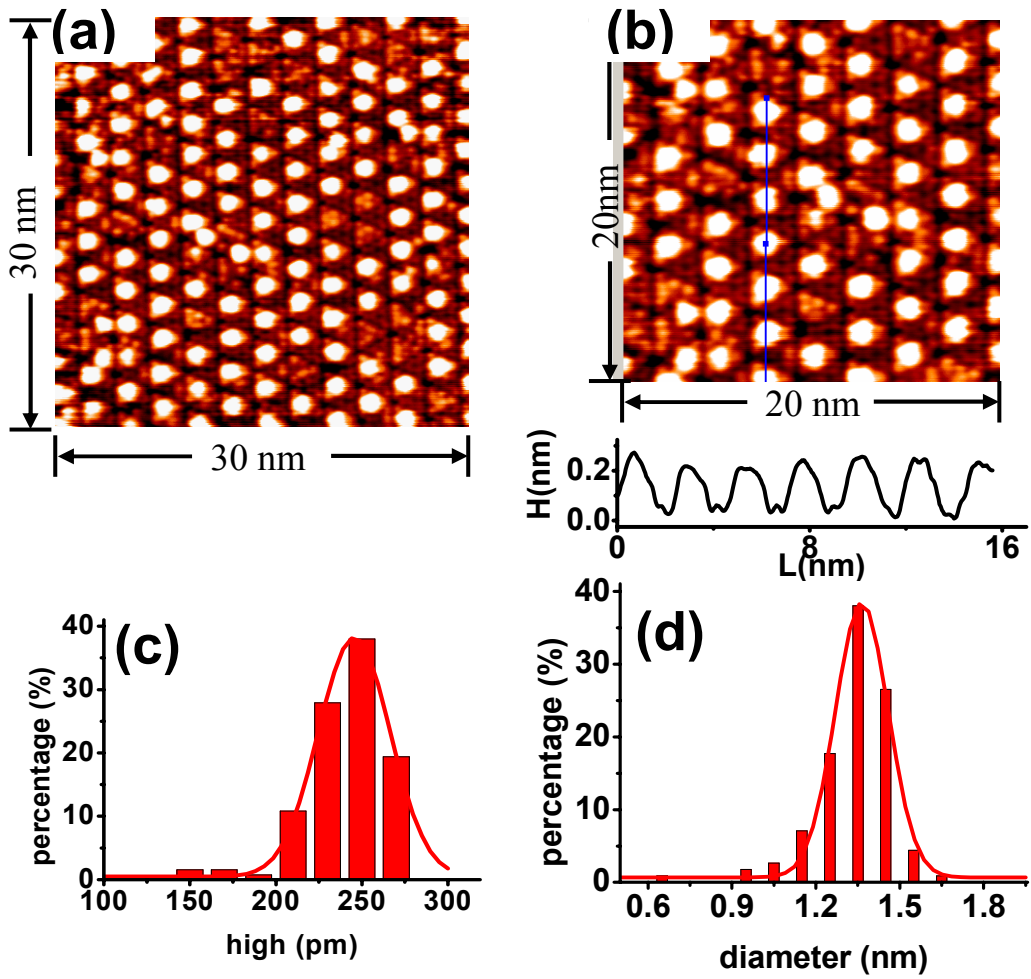

FIG3

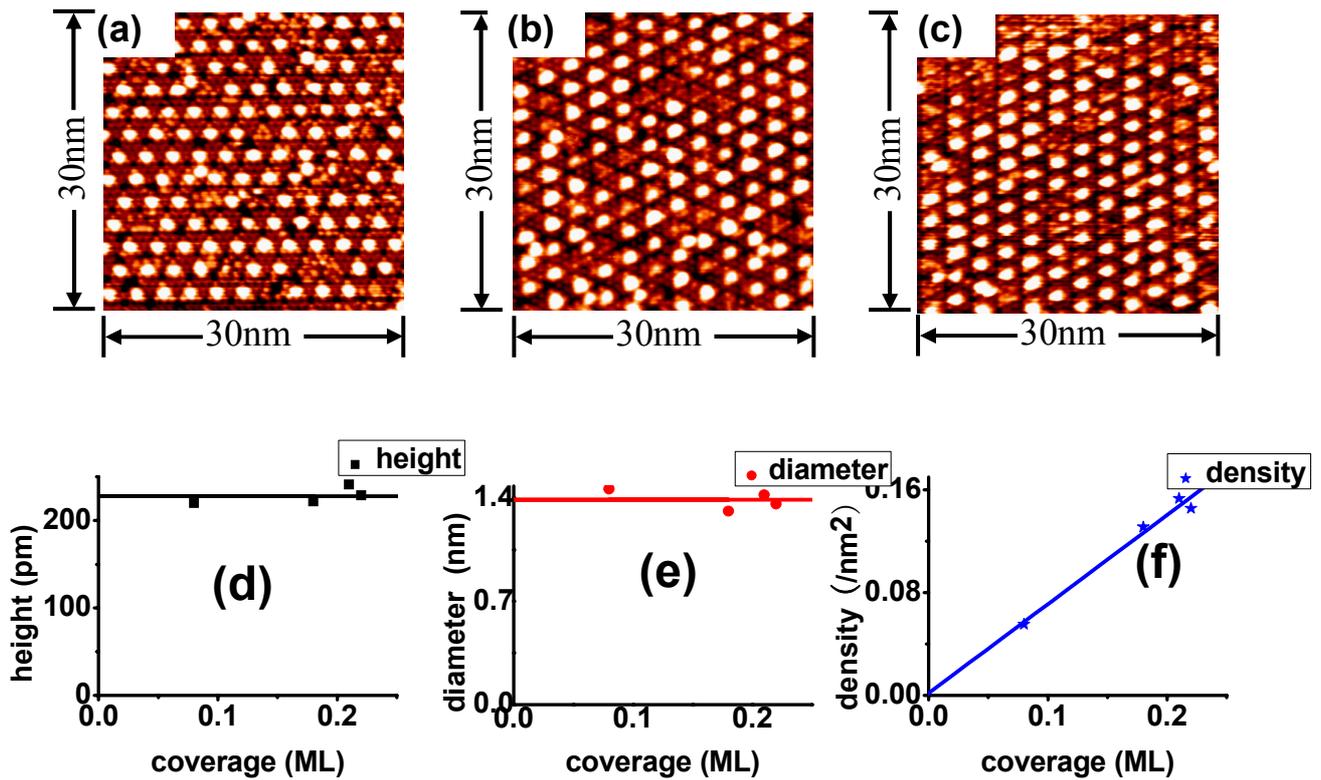

FIG. 4. (a), (b), and (c) are STM images of Mn clusters deposited on Si(111)-7×7 with different Mn dose. Every scanning area is 30×30 nm$^2$. The sample bias and tunneling current are -2.0V and 0.27nA respectively. (d), (e), and (f) show the relationship between the heights, diameters and densities of the Mn nanoclusters and the deposition coverage. The black and red line in (d), (e) is the mean of the height and the diameter of the clusters. Fig (f) shows the linear relationship between the areal densities of Mn nanoclusters and the deposition coverage.